%% file: main.tex
\documentclass[camera,letterpaper,nomarginnotes,nonarrowgutter]{jpaper}

\usepackage{footnote}
\makesavenoteenv{tabular}
\makesavenoteenv{table}

\DeclareMathAlphabet{\mathcal}{OMS}{cmsy}{m}{n}
\usepackage{array}
\usepackage{eso-pic}

\usepackage{fancyhdr}
\usepackage{datetime}
\usepackage[bookmarks=true,breaklinks=true,letterpaper=true,colorlinks,linkcolor=black,citecolor=black,urlcolor=black]{hyperref}
\usepackage{multirow}
\usepackage{multicol}
\usepackage{float}
\usepackage[nomessages]{fp}
\usepackage{dblfloatfix}    %
\usepackage{bm}
\usepackage{tikz}
\usepackage{atbegshi}
\usepackage{xcolor}
\usepackage[font={small,bf}]{subcaption}
\usepackage[linesnumbered,ruled]{algorithm2e}
\usepackage{balance}
\usepackage{siunitx}
\usepackage[resetlabels]{multibib}
\usepackage{hyperref}

\newcites{sidebar}{Sidebar: References}

\def\BibTeX{{\rm B\kern-.05em{\sc i\kern-.025em b}\kern-.08em
    T\kern-.1667em\lower.7ex\hbox{E}\kern-.125emX}}

\usepackage[font={small,bf}, figurename=Fig.]{caption}
\setlist[itemize]{leftmargin=*}

\usepackage{listings}
\usepackage{amsmath}

\usepackage{siunitx} 
\usepackage{glossaries}
\usepackage{makecell} %

 \setlist[itemize]{noitemsep,itemsep=0pt,parsep=0pt,topsep=0pt,partopsep=0pt,leftmargin=1.5em}
 \setlist[enumerate]{noitemsep,itemsep=0pt,parsep=0pt,topsep=0pt,partopsep=0pt,leftmargin=1.5em}

\usepackage{titlesec}
\titlespacing\section{0pt}{3pt plus 1pt minus 1pt}{2pt plus 1pt minus 1pt}
\titlespacing\subsection{0pt}{3pt plus 1pt minus 1pt}{2pt plus 1pt minus 1pt}
\titlespacing\subsubsection{0pt}{3pt plus 1pt minus 1pt}{2pt plus 1pt minus 1pt}

\input{macros}
 \renewcommand{\agyl}[1]{\agycomment{#1}}
 \renewcommand{\oml}[1]{\onurcomment{#1}}
 \renewcommand{\hluol}[1]{\hluocomment{#1}}

\input{glossary}
\makeatletter
\newcommand*{\textoverline}[1]{$\overline{\raisebox{0pt}[0.85\height]{#1}}\m@th$}
\makeatother

\newcommand{\affilETH}[0]{\textsuperscript{\S}}

\title{{RowPress Vulnerability in Modern DRAM Chips}}

\author{\vspace{-22pt}\\%
\fontsize{11}{12}\selectfont%
{Haocong Luo\affilETH{}}\quad%
{Ataberk Olgun\affilETH{}}\quad%
{A. Giray Ya\u{g}l{\i}k\c{c}{\i}\affilETH{}}\quad%
{Yahya Can Tu\u{g}rul\affilETH{}}\quad%
{Steve Rhyner\affilETH{}}\quad%
\vspace{-1pt}\\%
\fontsize{11}{12}\selectfont
{Meryem Banu Cavlak\affilETH{}}\quad%
{Jo{\"e}l Lindegger\affilETH{}}\quad%
{Mohammad Sadrosadati\affilETH{}}\quad%
{Onur Mutlu\affilETH{}}%
\vspace{0pt}\\%
{\fontsize{10}{11}\selectfont
\affilETH\emph{ETH Z{\"u}rich}%
}
\vspace{-6pt}}

\begin{document}

\bstctlcite{IEEEexample:BSTcontrol}
\bstctlcite[@auxoutsidebar]{IEEEexample:BSTcontrol}

\setlength{\footskip}{35pt} %

\fancyfoot[C]{\thepage}    %
\renewcommand{\headrulewidth}{0pt}
\renewcommand{\footrulewidth}{0pt}

\fancypagestyle{plain}{
  \fancyhf{}
  \fancyhead[C]{}     %
  \renewcommand{\footrulewidth}{0pt} 
  \renewcommand{\headrulewidth}{0pt}
}

\maketitle
\thispagestyle{fancy}

\begin{abstract}
    {Memory isolation is a critical property for system reliability, security, and safety. We demonstrate RowPress, a DRAM read disturbance phenomenon different from the well-known RowHammer. {RowPress} induces bitflips by keeping a DRAM row open for a long period of time instead of repeatedly opening and closing the row. We experimentally characterize RowPress bitflips, showing {their} widespread existence in commodity off-the-shelf DDR4 DRAM chips. We demonstrate RowPress bitflips in a real system that already has RowHammer protection, and propose effective mitigation techniques that {protect DRAM against} both RowHammer and RowPress.}
\end{abstract}

\input{01_motivation}

\input{02_rowpress}

\input{03_experimental_characterization}
\input{04_real_system}
\input{05_mitigation}

\input{06_significance}

\section{Author Biography}
\textbf{Haocong Luo} is a Ph.D student at the SAFARI research group, ETH Zurich, advised by Prof. Onur Mutlu. His current research interest is in computer architecture, focusing on the performance and robustness of memory systems. He obtained MSc. in Computer Science from ETH Zurich, and BEng. in Computer Science from ShanghaiTech University. \textbf{Contact email address:} \href{mailto:richardluo723@gmail.com}{richardluo723@gmail.com}

\textbf{Ataberk Olgun} is a Ph.D student at ETH Zürich, Zürich, Switzerland. Olgun received his master's degree in Computer Engineering from TOBB University of Economics and Technology. His research interests lie primarily in the area of Computer Architecture. In particular, his research interests include power-efficient and high-performance memory systems, and hardware security. \textbf{Contact email address:} \href{mailto:olgunataberk@gmail.com}{olgunataberk@gmail.com}

\textbf{Yahya Can Tugrul} received the bachelor’s degree in electrics and electronics engineering from Middle East Technical University, Ankara, Türkiye, in 2020. He is currently pursuing the master’s degree in computer engineering with the TOBB University of Economics and Technology, Ankara. He is a Scientific Assistant with ETH Zurich, Zürich, Switzerland. His research interests include secure and reliable DRAM architectures.
\textbf{Contact email address:} \href{mailto:yahyacantugrul@gmail.com}{yahyacantugrul@gmail.com}

\textbf{Steve Rhyner} is a Computer Science MSc student at ETH Zurich. He holds a Bachelor of Science ETH in Computer Science degree from ETH Zurich. His research interests span Machine Learning, Artificial Intelligence, Data Science, and their intersection with Computer Architecture. He is a member of the SAFARI research group.
\textbf{Contact email address:} \href{mailto:steverhyner7@gmail.com}{steverhyner7@gmail.com}

\textbf{Abdullah Giray Ya\u{g}l{\i}k\c{c}{\i}} recently defended his PhD thesis, advised by Prof. Onur Mutlu. His current broader research interests are in computer architecture and hardware security, with a special focus on DRAM robustness and performance. In particular, his PhD research focuses on understanding and solving DRAM read disturbance vulnerability. Giray has published several works on this topic in major venues such as HPCA, MICRO, ISCA, USENIX Security, DSN, and SIGMETRICS. Among these works, BlockHammer was named as a finalist by Intel in 2022 for the Intel Hardware Security Academic Award, Svärd received the first place in the ACM SRC at PACT 2023, and his dissertation was selected as one of the five finalists in the HOST 2024 PhD dissertation competition. Giray’s RowHammer research is in part supported by Google Security and Privacy Research Award and the Microsoft Swiss Joint Research Center.
\textbf{Contact email address:} \href{mailto:giray.yaglikci@safari.ethz.ch}{giray.yaglikci@safari.ethz.ch}

\textbf{Joël Lindegger} is a Ph.D student at the SAFARI research group, ETH Zürich. He received his master's degree in Computer Science from ETH Zürich. His research interests lie primarily in the area of accelerating genome analysis with software and hardware approaches.
\textbf{Contact email address:} \href{mailto:jmlindegger@gmail.com}{jmlindegger@gmail.com} 

\textbf{Meryem Banu Cavlak} received her bachelor’s degree in Computer Engineering from Bilkent University, Ankara, Türkiye in 2021, and master’s degree in Electrical Engineering and Information Technology from ETH Zürich, Zürich, Switzerland in 2024. Her research interests include bioinformatics and computer architecture.
\textbf{Contact email address:} \href{mailto:mbanucavlak@gmail.com}{mbanucavlak@gmail.com}  

\textbf{Mohammad Sadrosadati} is a senior researcher and lecturer at SAFARI Research Group, ETH Zurich, working under the supervision of Prof. Onur Mutlu. His research interests are in near-data processing, memory/storage systems, heterogeneous computing, and interconnection networks. He received the B.Sc., M.Sc., and Ph.D. degrees in Computer Engineering from Sharif University of Technology, Tehran, Iran, in 2012, 2014, and 2019, respectively.
\textbf{Contact email address:} \href{mailto:m.sadr89@gmail.com}{m.sadr89@gmail.com} 

\textbf{Onur Mutlu} is a Professor of Computer Science at ETH Zurich. He is also a faculty member at Carnegie Mellon University, where he previously held the Strecker Early Career Professorship. His research interests are in computer architecture, systems, hardware security, bioinformatics. A variety of techniques he, along with his group and collaborators, has invented over the years have influenced industry and are employed in commercial microprocessors and memory/storage systems. He started the Computer Architecture Group at Microsoft Research (2006-2009), and held various product and research positions at Intel Corporation, AMD, VMware, and Google.  He is an ACM Fellow, IEEE Fellow, and an elected member of the Academy of Europe. His computer architecture and digital design course lectures are freely available (\url{https://www.youtube.com/OnurMutluLectures}). His research group makes a wide variety of software and hardware artifacts freely available online (\url{https://safari.ethz.ch/}). \textbf{Contact email address:} \href{mailto:omutlu@gmail.com}{omutlu@gmail.com}

\section*{{Acknowledgements}}
{We thank the anonymous reviewers of ISCA 2023 and IEEE Micro Top Picks 2024 for feedback. We thank {the} SAFARI Research Group members for {valuable} feedback and the stimulating intellectual environment they provide. We acknowledge the generous gift funding provided by our industrial partners ({especially} Google, Huawei, Intel, Microsoft, VMware), which has been instrumental in enabling the research we have been conducting on read disturbance in DRAM in particular and memory systems in general. This work was in part supported by the {a Google Security and Privacy Research Award and the Microsoft Swiss Joint Research Center}.}

\bibliographystyle{IEEEtran}
\bibliography{refs}

\input{07_related_works}

\end{document}

%% file: macros.tex
\definecolor{amber}{rgb}{1.0, 0.49, 0.0}
\definecolor{awesome}{rgb}{1.0, 0.13, 0.32}
\definecolor{dollarbill}{rgb}{0.52,0.73,0.4}
\definecolor{moegi}{rgb}{0.357, 0.537, 0.188}
\definecolor{burgundy}{rgb}{0.5, 0.0, 0.13}
\definecolor{ballblue}{rgb}{0.13, 0.67, 0.8}
\definecolor{ups-truck}{rgb}{0.53, 0.28, 0.21}
\definecolor{airforceblue}{rgb}{0.36, 0.54, 0.66}
\definecolor{cadmiumgreen}{rgb}{0.0, 0.42, 0.24}
\definecolor{darkcyan}{rgb}{0.0, 0.55, 0.55}
\definecolor{caribbeangreen}{rgb}{0.0, 0.8, 0.6}
\definecolor{flamingopink}{rgb}{0.99, 0.56, 0.67}
\definecolor{jazzberryjam}{rgb}{0.65, 0.04, 0.37}
\definecolor{mediumpersianblue}{rgb}{0.0, 0.4, 0.65}
\definecolor{coolblack}{rgb}{0.0, 0.18, 0.39}
\definecolor{bleudefrance}{rgb}{0.19, 0.55, 0.91}
\definecolor{ao}{rgb}{0.0, 0.0, 1.0}
\definecolor{babyblueeyes}{rgb}{0.63, 0.79, 0.95}
\definecolor{antiquefuchsia}{rgb}{0.57, 0.36, 0.51}
\definecolor{whitesmoke}{rgb}{0.96, 0.96, 0.96}

\newcommand{\om}[1]{{#1}}

\newcommand{\squishlist}{
 \begin{list}{$\circ$}
  { \setlength{\itemsep}{0pt}
     \setlength{\parsep}{0pt}
     \setlength{\topsep}{0pt}
     \setlength{\partopsep}{0pt}
     \setlength{\leftmargin}{1em}
     \setlength{\labelwidth}{1em}
     \setlength{\labelsep}{0.5em} } }

\newcommand{\squishsublist}{
\begin{list}{$\rightarrow$}
 { \setlength{\itemsep}{0pt}
    \setlength{\parsep}{0pt}
    \setlength{\topsep}{-10em}
    \setlength{\partopsep}{-3pt}
    \setlength{\leftmargin}{1em}
    \setlength{\labelwidth}{1em}
    \setlength{\labelsep}{0.5em} } }

\newcommand{\squishend}{
  \end{list}  }

\newcounter{obs}
\newcounter{take}
\newcommand\take[1]{%
   \refstepcounter{take}
  \vspace{0.3em}
  \noindent
  \begin{tabular}{|p{0.95\linewidth}|}
       \hline
       \textbf{{Takeaway \thetake}.} \emph{{#1}}\\
       \hline 
  \end{tabular}
}

\newif\ifcameraready
\camerareadytrue
\camerareadyfalse

\newif\ifrevision
\revisionfalse

\newif\ifsubmission
\submissionfalse

\newif\ifonurready
\onurreadytrue

\ifcameraready

    \newcommand{\om}[1]{{#1}}

    \newcommand{\atbc}[1]{}
    \newcommand{\loiscomment}[1]{}
    \newcommand{\onurcomment}[1]{}
    \newcommand{\agycomment}[1]{}
    \newcommand{\hluocomment}[1]{}
    \newcommand{\jktodo}[1]{}
    \newcommand{\mptodo}[1]{}
    
    \newcommand{\agyl}[1]{}
    \newcommand{\oml}[1]{}
    \newcommand{\hluol}[1]{}
    
    \newcommand{\ignore}[1]{}

\else
\ifonurready 

    \newcommand{\atbc}[1]{\textcolor{blue}{\textbf{[@atb: #1]}}}
    \newcommand{\loiscomment}[1]{\noindent{\color{blue}{\bf \fbox{Lois: }}{\it#1}}}
    \newcommand{\onurcomment}[1]{\noindent{\color{blue}{\bf \fbox{ONUR: }}{\it#1}}}
    \newcommand{\agycomment}[1]{\textcolor{blue}{\textbf{[@gy:~{\it#1}]}}}

    \newcommand{\hluocomment}[1]{\textcolor{blue}{\textit{[@hluo: #1]}}}

    \newcommand{\agyl}[1]{\todo[size=\scriptsize,color=babyblueeyes]{\textbf{{@gy:}} #1}}
    \newcommand{\oml}[1]{\todo[size=\scriptsize,color=babyblueeyes]{\textbf{{onur:}} #1}}
    
    \newcommand{\hluol}[1]{\todo[size=\scriptsize,color=babyblueeyes]{\textbf{{@hluo:}} #1}}
    
    \newcommand{\jktodo}[1]{\textcolor{blue}{\textbf{[JKTODO:}#1\textbf{]}}}
    \newcommand{\mptodo}[1]{\textbf{\scriptsize \fcolorbox{black}{red}{\color{white}{TODO}}}\textcolor{red}{\emph{#1}}}

\else
    \ifrevision

      \newcommand{\agyl}[1]{\todo[size=\small,color=babyblueeyes]{\textbf{{@gy:}} #1}}
      \newcommand{\hluol}[1]{\todo[size=\scriptsize,color=babyblueeyes]{\textbf{{@hluo:}} #1}}

      \newcommand{\om}[1]{\textcolor{coolblack}{#1}}

      \newcommand{\agycomment}[1]{}
      \newcommand{\loiscomment}[1]{}
      \newcommand{\onurcomment}[1]{}
      \newcommand{\hluocomment}[1]{}
      \newcommand{\jktodo}[1]{}
      \newcommand{\mptodo}[1]{}
      \newcommand{\atbc}[1]{}
      \newcommand{\ignore}[1]{}
      
    \else

      \ifsubmission   %

        \newcommand{\om}[1]{#1}

        \newcommand{\agycomment}[1]{}
        \newcommand{\loiscomment}[1]{}
        \newcommand{\onurcomment}[1]{}
        \newcommand{\hluocomment}[1]{}
        \newcommand{\atbc}[1]{}
        \newcommand{\ignore}[1]{}
        \newcommand{\jktodo}[1]{}
        \newcommand{\mptodo}[1]{}
        
        \newcommand{\agyl}[1]{}

    \else           %

        \newcommand{\jktodo}[1]{\textcolor{mediumpersianblue}{\textbf{[JKTODO:}#1\textbf{]}}}
        \newcommand{\mptodo}[1]{\textbf{\scriptsize \fcolorbox{black}{red}{\color{white}{TODO}}}\textcolor{red}{\emph{#1}}}

        \newcommand{\om}[1]{\textcolor{jazzberryjam}{#1}}

        \newcommand{\atbc}[1]{\textcolor{ups-truck}{\textbf{[@atb: #1]}}}
        
        \newcommand{\loiscomment}[1]{\noindent{\color{red}{\bf \fbox{Lois: }}{\it#1}}}
        \newcommand{\onurcomment}[1]{\noindent{\color{red}{\bf \fbox{ONUR: }}{\it#1}}}
        \newcommand{\agycomment}[1]{\textcolor{awesome}{\textbf{[@gy: {\it#1}]}}}
        \newcommand{\hluocomment}[1]{\textcolor{moegi}{\textit{[@hluo: #1]}}}
        \newcommand{\ignore}[1]{}
        \newcommand{\agyl}[1]{\todo[size=\scriptsize,color=babyblueeyes]{\textbf{{@gy:}} #1}}
        \newcommand{\oml}[1]{\todo[size=\scriptsize,color=babyblueeyes]{\textbf{{onur:}} #1}}
        \newcommand{\hluol}[1]{\todo[size=\scriptsize,color=babyblueeyes]{\textbf{{@hluo:}} #1}}
    \fi
  \fi
\fi
\fi

\newcommand{\tras}{t_{RAS}}
\newcommand{\trp}{t_{RP}}

\newacronym{hcfirst}{\textit{HC\textsubscript{first}}}{the minimum hammer count {value} at which the first bit error is observed}
\newacronym{ber}{$BER$}{the number of bit flips in a DRAM row, refer{red} to as bit error rate}
\newacronym{wcdp}{$WCDP$}{worst-case data pattern}

\newacronym{taggon}{$t_{AggOn}$}{the time that an aggressor row stays active, {i.e., aggressor row's on-time}}
\newacronym{taggoff}{$t_{AggOff}$}{the time that {the bank} stays precharged, {i.e., aggressor row's off-time}}
\newacronym{tras}{$\tras{}$}{the minimum time that a row should stay active before a precharge command is issued {to the bank}}
\newacronym{trp}{$\trp{}$}{the minimum time a precharge command needs to complete before a row is activated {in the same bank}}
\newacronym{iqr}{$IQR$}{interquartile range}
\newacronym{cv}{$CV$}{the coefficient of variation}
\newacronym{hc}{$HC$}{hammer count}
\newacronym{rfm}{$RFM$}{refresh management}

\newcommand*\DRAMCMD[1]{\texttt{#1}}
\newcommand*\DRAMTIMING[1]{t\textsubscript{#1}}
\newcommand*\nCHIPS{164}

\glsdisablehyper

\newacronym{vdd}{$V_{DD}$}{supply voltage}
\newacronym{vpp}{$V_{PP}$}{wordline voltage}
\newacronym{vwl}{$V_{PP}$}{wordline voltage}
\newacronym{vgs}{$V_{GS}$}{gate-to-source voltage}
\newacronym{vth}{$V_{TH}$}{the voltage threshold that the bitline voltage should exceed for the activation to be reliably completed}
\newacronym{gnd}{$GND$}{ground}

\newacronym{acmin}{$AC_{min}$}{the minimum number of total aggressor row activations to cause at least one bitflip}
\newacronym{ac}{$AC$}{activation count}
\newacronym{rblast}{$r_{Blast}$}{blast radius}

\newacronym{trcd}{\DRAMTIMING{RCD}}{
{the minimum time between opening a row with an \DRAMCMD{ACT} command and accessing the row buffer}
}

\newacronym{trefw}{\DRAMTIMING{REFW}}{the maximum time window between two refresh operations that target {the same} row}

%% file: 01_motivation.tex
\section{Motivation}
\label{sec:motivation}
{To ensure system {robustness (i.e., reliability, security, and safety)}, it is critical to} maintain memory isolation: accessing a memory address should not cause unintended side-effects on data stored in other addresses. {Unfortunately}, with aggressive technology node scaling, {dynamic random access memory (DRAM), the prevalent {main} memory technology}, suffers from increased {{\emph{read disturbance}}: accessing (reading) a DRAM cell disturbs the operational characteristics (e.g., stored charge) of other physically close DRAM cells.} 

RowHammer{~\cite{kim2014flipping}} is an example read-disturb phenomenon where \emph{repeatedly} opening and closing (i.e., hammering) a DRAM row (called aggressor row) \emph{many times} (e.g., tens of thousands times) can {cause \emph{bitflips}} in physically nearby rows (called victim rows). RowHammer {is} a critical security vulnerability as attackers can {induce and exploit the bitflips to take over a system or leak private or security-critical data{~\cite{mutlu2019rowhammer}}.} To ensure {robust} operation in {modern and future} DRAM-based systems, it is critical to develop a {rigorous} understanding of {read disturbance} {effects like RowHammer}.

%% file: 02_rowpress.tex
\section{RowPress}

\begin{figure}[h]
    \centering
    \includegraphics[width=\linewidth]{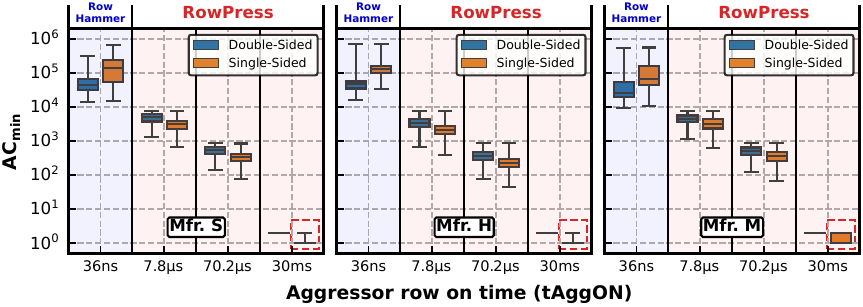}
    \caption{{$AC_{min}$ {distributions} of conventional RowHammer (RH) and three representative cases of RowPress (RP) at $80^{\circ}C$ across 164 DDR4 chips {from manufacturers S, H, and M}}.}
    \label{fig:hcf_intro}
\end{figure}

{Our ISCA 2023 paper~\cite{luo2023rowpress}} experimentally demonstrates {another widespread} {read-disturb phenomenon} {different from RowHammer},
\emph{RowPress}, in {real commodity off-the-shelf} DDR4 DRAM chips. We show that keeping a DRAM row {(i.e., aggressor row)} open for a {long period of time} (i.e., {having a long aggressor row on time, \DRAMTIMING{AggON}}) disturbs physically nearby DRAM rows. {Doing so} induces bitflips {in the victim row} \emph{without} {requiring {(tens of)} thousands of activations to the} aggressor row like RowHammer. 
{To illustrate this, Figure~\ref{fig:hcf_intro} shows the distribution of \glsdesc{acmin} ($AC_{min}$) of RowHammer and RowPress. Each box-and-whiskers represents the $AC_{min}$ distribution (y-axis) when the aggressor DRAM row is open for \DRAMTIMING{AggON} (x-axis) we measure {in \nCHIPS{}} {commodity off-the-shelf} DRAM chips {from} all three major DRAM manufacturers (i.e., Mfr. S, H, and M, corresponding to Samsung, SK Hynix, Micron; see Table~\ref{tab:dram_chip_list}). {The box represents the the first (lower edge) and third (upper edge) quartile of the measured $AC_{min}$ values, and the whiskers show the minimum and maximum values.} The DRAM temperature is set {to} \SI{80}{\celsius} {(still within the ``Normal Operating Temperature Range'' as defined by the JEDEC DDR4 standard~\cite{jedec2017ddr4})}.}

{The two leftmost boxes in each {plot} {show} the distribution of \gls{acmin} {for the conventional single-sided ({orange}) and double-sided (blue) RowHammer {patterns}, where} the aggressor row is open for the minimum amount of time {($\DRAMTIMING{AggON} = \DRAMTIMING{RAS} = \SI{36}{\nano\second}$)} allowed by the DRAM specification~\cite{jedec2017ddr4}, as done in conventional RowHammer attacks. {We observe that as \DRAMTIMING{AggON} increases, compared to the most effective RowHammer pattern, the most effective RowPress pattern reduces \gls{acmin}}}
{1)~by {$17.6\times$} on average (up to {$40.7\times$}) when \DRAMTIMING{AggON} {is as large as} the refresh interval (\SI{7.8}{\micro\second}),
{2)~by {$159.4\times$} on average (up to {$363.8\times$}) when \DRAMTIMING{AggON} is \SI{70.2}{\micro\second}, the maximum {allowed} \DRAMTIMING{AggON} according to the current JEDEC standard~\cite{jedec2017ddr4},} and {3)~in extreme cases, down to {\emph{only one}} activation (e.g., when \DRAMTIMING{AggON} $=$ \SI{30}{\milli\second}, {highlighted by red {boxes})}.}}

To our knowledge, {our ISCA 2023 paper is} the first work to {experimentally} {demonstrate} {the RowPress phenomenon} and {its widespread existence in {real} {commodity off-the-shelf} DDR4 DRAM chips from all three major {DRAM} manufacturers. {In this article,} we 1) provide an extensive and rigorous real-device characterization of RowPress, demonstrating its unique characteristics that {are} different from RowHammer, 2) {as a proof of concept}{, develop a \emph{user-level} program that leverages RowPress to induce bitflips on a real system {with DRAM} that already has RowHammer protection}, and 3) propose {and evaluate a systematic} methodology to adapt existing RowHammer mitigation {techniques} to \emph{also} mitigate RowPress. {Our results} suggest that DRAM-based systems need to take RowPress into account to {maintain the fundamental security{/safety/reliability} property of memory isolation {and achieve robust operation}}. Based on our findings, we discuss and evaluate the implications of RowPress {on} existing {{read-disturb} mitigation} mechanisms that consider \emph{only} RowHammer. {{To enable 1) reproduction and replication of our results, and 2) further research on RowPress, we open-source all our infrastructure, test programs, and data at \url{https://github.com/CMU-SAFARI/RowPress}.}} {Extended results of our work and detailed analyses can be found in the arXiv version of our paper~\cite{rowpress-arxiv}.}}

%% file: 03_experimental_characterization.tex
\section{Real {DRAM Chip} Characterization}
\subsection{{Characterization Methodology}}

\begin{figure}[h]
    \centering
    \includegraphics[width=0.95\linewidth]{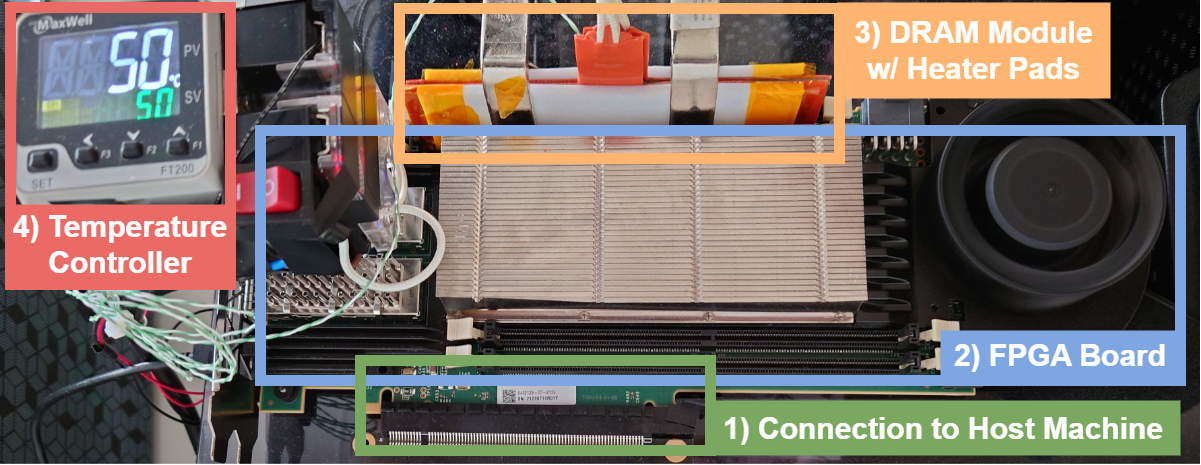}
    \caption{Our DDR4 DRAM testing infrastructure.}
    \label{fig:infra_photo}
\end{figure}

\noindent {\textbf{Infrastructure.}} We characterize {commodity} DDR4 DRAM chips using an FPGA-based DRAM testing infrastructure based on DRAM Bender~\cite{olgun2022drambender} {that enables us fine-grained control of the DRAM commands, timings, and temperature of the DRAM chips.} {Figure~\ref{fig:infra_photo} shows the four main components of our infrastructure: 1)~a host machine that generates the characterization {program} and {collects experiment results}, 2)~an FPGA development board that the memory module with the DRAM chips under characterization {are} connected to, programmed with DRAM Bender~\cite{olgun2022drambender} that executes the characterization programs, 3)~a thermocouple temperature sensor and a pair of heater pads pressed against the DRAM chips, and 4)~a PID temperature controller that controls the heaters and maintains the temperature at the desired level.}

\noindent {\textbf{DRAM Chips Characterized.}}  We characterize 164 real DDR4 DRAM chips from 21 {commodity off-the-shelf} DRAM modules ({shown in Table~\ref{tab:dram_chip_list}}). To {{demonstrate} that RowPress is intrinsic to the DRAM technology {and is a widespread phenomenon {across manufacturers},}} we test a variety of DRAM chips spanning different die densities and die revisions from all three major DRAM chip manufacturers (Mfr. S, H, and M). 
 
\input{03a_dram_table}

\noindent {\textbf{Key Metric.}} {We quantify DRAM's vulnerability to read disturbance by {measuring} \glsdesc{acmin} (\gls{acmin}). {A lower \gls{acmin} value means the DRAM is \emph{more vulnerable} to read disturbance because \emph{fewer} aggressor row activations are needed to induce bitflips.} For every DRAM module, we characterize \gls{acmin} for {3072} DRAM rows (the first, the middle, and the last {1024} rows in bank 1) and report the minimum \gls{acmin} value across five repetitions. {To avoid bitflips due to retention failures}, we bound the duration of the \gls{acmin} characterization routine \emph{strictly within} the \SI{64}{\milli\second} refresh window of the JEDEC DDR4 standard~\cite{jedec2017ddr4}.}

\subsection{Key Characterization Results}
\begin{figure}[h]
    \centering
    \includegraphics[width=\linewidth]{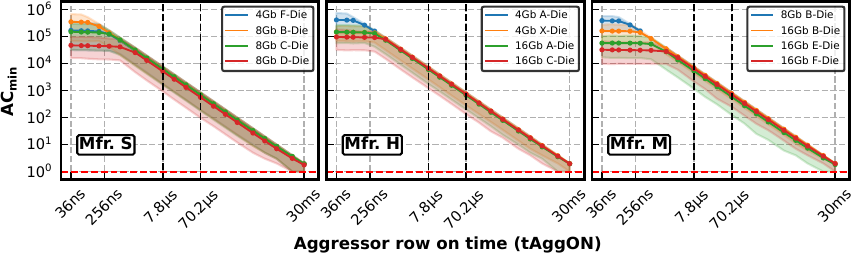}
    \caption{How \gls{acmin} changes as \DRAMTIMING{AggON} {increases}.}
    \label{fig:acmin_taggon}
\end{figure}

First, we characterize how RowPress amplifies DRAM's vulnerability to read disturbance by analyzing how \gls{acmin} changes as the aggressor row on time, \DRAMTIMING{AggON}, increases. We use a single-sided RowPress access pattern involving only one aggressor row, a fixed checkerboard data pattern {(i.e., the aggressor row is initialized with \texttt{0xAA}, and the victim rows are initialized with \texttt{0x55})}, and keep the DRAM temperature at \SI{50}{\celsius} (more access patterns, data patterns, and temperatures are examined in our ISCA 2023 paper~\cite{luo2023rowpress} {and its extended version~\cite{rowpress-arxiv}).} {Figure~\ref{fig:acmin_taggon} shows the \gls{acmin} {distribution} {(y-axis)} {of different die revisions for all three major DRAM manufacturers} as we sweep \DRAMTIMING{AggON} {(x-axis)} from \SI{36}{\nano\second} to \SI{30}{\milli\second} in log-log scale. {Note that when \DRAMTIMING{AggON} is \SI{36}{\nano\second} (i.e., the left-most data points), the access pattern we characterize is identical to a conventional single-sided RowHammer pattern.} {For each manufacturer ({i.e., each plot}), we group the data based on the die revision {(different colors)} and then aggregate the \gls{acmin} values from all the rows we test}. Each data point shows the mean \gls{acmin} value and the error band shows the {minimum and maximum of \gls{acmin} values} {across all tested rows}. We highlight the \DRAMTIMING{AggON} values of \SI{7.8}{\micro\second} (\DRAMTIMING{REFI}) and \SI{70.2}{\micro\second} ($9\times$\DRAMTIMING{REFI}) on the x-axis, as they are the two potential upper bounds of \DRAMTIMING{AggON}, as {dictated by} the JEDEC DDR4 standard~\cite{jedec2017ddr4}. {We mark \gls{acmin} $=$ 1 {($10^{0}$)} on the y-axis {with dashed red lines}.}}

 We observe that across DRAM chips from all three manufacturers, 1) \gls{acmin} reduces by 21$\times$ (190$\times$) on average when \DRAMTIMING{AggON} increases from \SI{36}{\nano\second} to \SI{7.8}{\micro\second} (\SI{70.2}{\micro\second}){, and 2) {under} extreme conditions (i.e., when \DRAMTIMING{AggON} $=$ \SI{30}{\milli\second}), \emph{only a single} aggressor row activation is enough to cause bitflips}. Our results also show that DRAM chips with presumably more advanced technology nodes are more vulnerable to RowPress.

\take{RowPress is a common read-disturb phenomenon {in DRAM chips} that {exacerbates} {DRAM's vulnerability to read disturbance}.}

Second, we compare RowPress bitflips induced with \gls{acmin} (\DRAMTIMING{AggON} $>$ \SI{36}{\nano\second}) to 1) RowHammer bitflips that are also induced at \gls{acmin} (\DRAMTIMING{AggON} $=$ \SI{36}{\nano\second}), and 2) retention failure bitflips induced with an accelerated retention failure profiling methodology {(i.e., we set the DRAM temperature to \SI{80}{\celsius}, initialize the DRAM rows with the checkerboard data pattern, and disable auto-refresh for four seconds)}. {Our results show that 1) an overwhelming majority of RowPress bitflips are different {from} those caused by RowHammer and retention failures, and 2) RowPress and RowHammer bitflips have \emph{opposite} directions. For example, 1) for \DRAMTIMING{AggON} $\geq$ \SI{7.8}{\micro\second}, on average, only less than $0.013\%$ of DRAM cells vulnerable to RowPress overlap with those vulnerable to RowHammer, and less than {$0.34\%$} overlap with retention failures, and 2) the majority of RowHammer bitflips in the DRAM chips we characterize are from {logical} ``0'' to {logical} ``1'' while the majority of RowPress bitflips are from {logical} ``1'' to {logical} ``0'' {(note that logical bit values do \emph{not} always correspond to physical charge levels stored in the DRAM cell~\cite{liu2013experimental})}. These results suggest that different failure mechanisms lead to RowPress and RowHammer bitflips.}

\take{RowPress has a different failure mechanism {from} RowHammer and data retention failures in DRAM. There is almost no overlap between RowPress, RowHammer, and data retention bitflips, and the directionality of RowHammer and RowPress bitflips show opposite trends.}

Third, we study the sensitivity of RowPress bitflips with respect to the change in temperature and access pattern. {We observe that at a higher temperature of \SI{80}{\celsius}, RowPress reduces \gls{acmin} even more compared to \SI{50}{\celsius}. {For example, when \DRAMTIMING{AggON} is {\SI{7.8}{\micro\second}}, the average \gls{acmin} at \SI{80}{\celsius} is only 0.55$\times$, 0.32$\times$, and 0.59$\times$ of that at \SI{50}{\celsius}, for Mfr. S, H, and M, respectively.}}

\take{RowPress gets significantly worse as temperature increases. This behavior is very different from how RowHammer bitflips change with temperature {(as shown in two prior works,~\cite{kim2014flipping} and~\cite{orosa2021deeper})}.}

\begin{figure}[h]
    \centering
    \includegraphics[width=\linewidth]{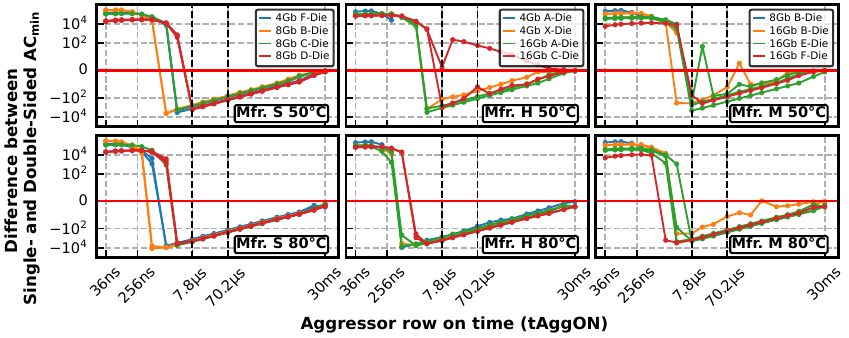}
    \caption{Single-sided \gls{acmin} minus double-sided \gls{acmin} at $50^{\circ}C$ (first row) and $80^{\circ}C$ (second row).}
    \label{fig:single-double}
\end{figure}

{Fourth,} we compare the \gls{acmin} values of the single-sided RowPress pattern involving only a single aggressor row to the double-sided RowPress pattern where two aggressor rows sandwiching a victim row are activated alternatingly. {Figure~\ref{fig:single-double} shows the difference between single- and double-sided \gls{acmin} (i.e., \gls{acmin}$(single)$ - \gls{acmin}$(double)$) at $50^{\circ}C$ (first row) and $80^{\circ}C$ (second row).} 

We observe that as \DRAMTIMING{AggON} increases, at both \SI{50}{\celsius} and \SI{80}{\celsius}, the \gls{acmin} of single-sided RowPress is initially larger than {that of} double-sided RowPress. However, as \DRAMTIMING{AggON} continues to increase, the \gls{acmin} of single-sided RowPress becomes \emph{smaller} than {that of} double-sided RowPress (i.e., single-sided RowPress becomes more effective at inducing bitflips because it needs \emph{less} aggressor row activations).

\take{RowPress behaves very differently from RowHammer as we change the access pattern from single-sided to double-sided. {As \DRAMTIMING{AggON} increases beyond a certain value, {RowPress} needs {fewer} aggressor row activations to induce bitflips with the single-sided pattern compared to {the double-sided pattern}.}}

{We do not fully understand the underlying reasons for the observed phenomena and the low-level failure mechanisms of RowPress. We call for more future work in this direction to build a more comprehensive understanding of RowPress and read disturbance in DRAM in general.}

{Our ISCA 2023 paper~\cite{luo2023rowpress} and its extended version~\cite{rowpress-arxiv}} provide {more comprehensive real-device characterization results and more detailed analyses, including the fraction of DRAM rows vulnerable to RowPress, RowPress bitflips' sensitivity to different data patterns, the minimum \DRAMTIMING{AggON} to induce at least one bitflip for a fixed aggressor row activation count, a more fine-grained temperature study, the effect of increased aggressor row off time (\DRAMTIMING{AggOFF}), and the repeatability of RowPress bitflips.}

%% file: 03a_dram_table.tex
\begin{scriptsize}
\begin{table}[h!]
  \centering
  \scriptsize
  \captionsetup{justification=centering, singlelinecheck=false, labelsep=colon}
  \caption{Tested DDR4 DRAM Chips.}
    \begin{tabular}{lcccccc}
        \toprule
            {{\bf Mfr.}} & \textbf{\#DIMMs} & {{\bf  \#Chips}}  & {{\bf Density}} & {{\bf Die Rev.}}& {{\bf Org.}}& {{\bf Date}}\\
        \midrule
\multirow{4}{4em}{Mfr. S (Samsung)} & 2 & 8  & 8Gb  & B   & x8  & 20-53  \\                           
  & 1 & 8  & 8Gb  & C   & x8  & N/A   \\                         
   & 3 & 8  & 8Gb  & D   & x8  & 21-10  \\                           
       & 2 & 8  & 4Gb  & F   & x8  & N/A   \\                         
        \midrule
\multirow{4}{4em}{Mfr. H (\mbox{SK Hynix})}     & 1 & 8  & 4Gb  & A   & x8  & 19-46  \\     
     & 1 & 8  & 4Gb  & X & x8  & N/A   \\               
     & 2 & 8  & 16Gb & A   & x8  & 20-51  \\                           
     & 2 & 8  & 16Gb & C   & x8  & 21-36  \\                           
     \midrule                  
\multirow{5}{4em}{Mfr. M (Micron)}     & 1 & 16 & 8Gb  & B   & x4  & N/A   \\       
    & 2 & 4  & 16Gb & B   & x16 & 21-26  \\                            
    & 1 & 16 & 16Gb & E   & x4  & 20-14  \\     
    & 2 & 4  & 16Gb & E   & x16 & 20-46  \\ 
    & 1 & 4  & 16Gb & F   & x16 & 21-50  \\                            
        \bottomrule
    \end{tabular}
    \label{tab:dram_chip_list}
\end{table}
\end{scriptsize}

%% file: 04_real_system.tex
\section{{Proof-of-Concept} Real System Demonstration}
\subsection{Methodology}
\begin{figure}[h]
    \centering
    \includegraphics[width=\linewidth]{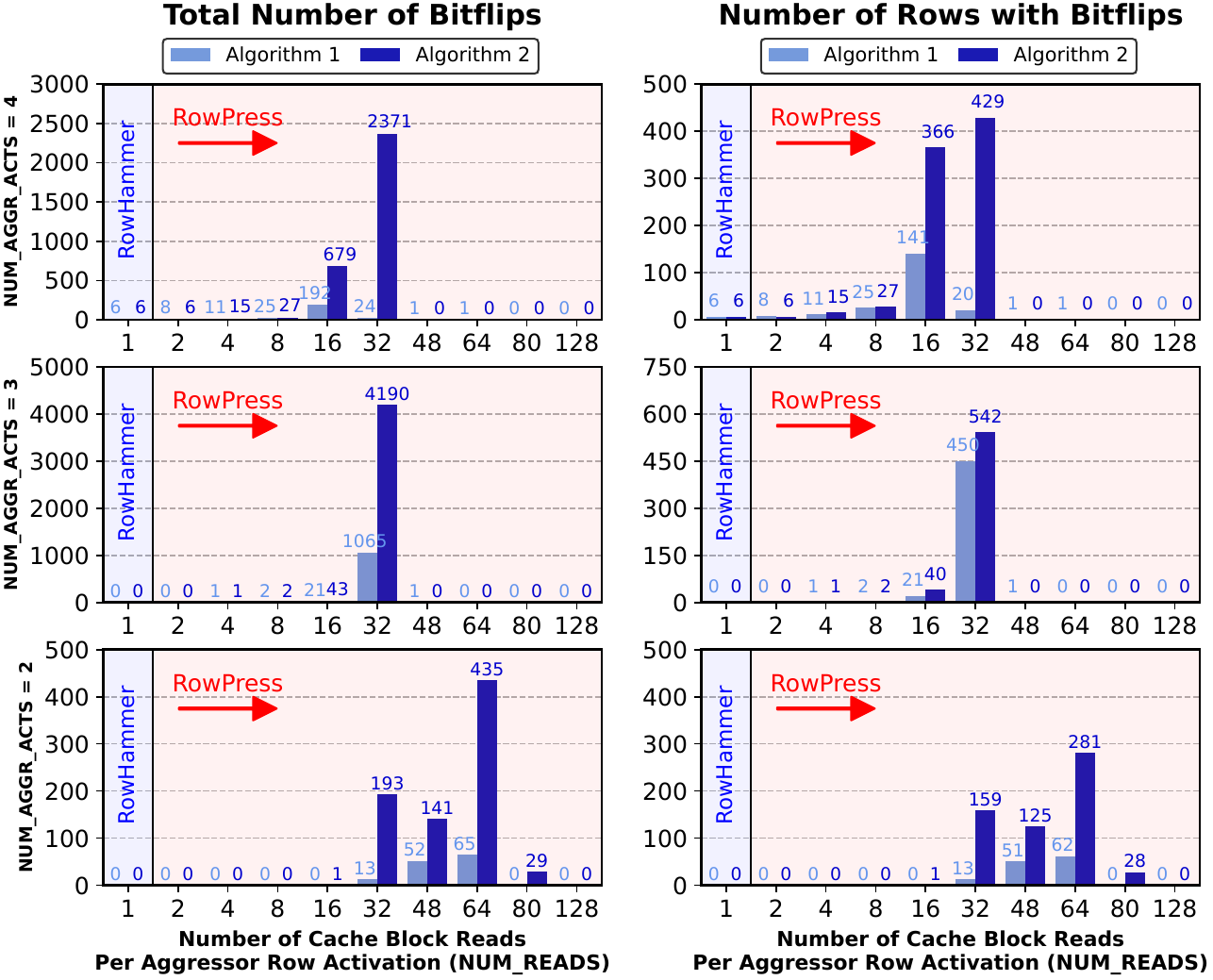}
    \caption{Number of RowHammer vs. RowPress bitflips (left) and number of rows with bitflips (right) we observe after running our {proof of concept test programs} with Algorithm 1 (blue bars) and Algorithm 2 (purple bars) {with four (top), three (middle), and two (bottom) activations per aggressor row per iteration}.}
    \label{fig:real}
\end{figure}

{We demonstrate that {an unprivileged} \emph{user-level} program can induce RowPress {bitflips} {in a real system that uses DDR4 main memory and already has RowHammer protection. The system {we evaluate} consists of an Intel Core i5-10400 {(Comet Lake)} processor and a DDR4 DRAM module from Mfr. S {with 8Gb C-Die DRAM chips}. The DRAM chips on this module has Target Row Refresh (TRR)~{\cite{hassan2021utrr}}, a widely adopted in-DRAM {RowHammer} mitigation mechanism~{\cite{frigo2020trrespass}} employed by DRAM manufacturers.}

\subsection{Demonstration Program}
{Algorithm~\ref{alg:rowpress-program} shows the key part of our RowPress demonstration program. The program repeatedly 1) accesses \emph{multiple} (i.e., \texttt{NUM\_READS}) cache blocks in \emph{the same aggressor DRAM row} (line 11 and 12) so that the memory controller keeps the aggressor row open for a longer period of time to better serve these accesses, and 2) flushes the accessed cache blocks from the cache (lines 14-16) so that future accesses will activate the aggressor row in DRAM. Note that when \texttt{NUM\_READS} is 1, the program is identical to a conventional RowHammer program.}
\renewcommand{\lstlistingname}{Algorithm} %
\begin{lstlisting}[caption={RowPress demonstration program.},captionpos=b,label={alg:rowpress-program}, basicstyle=\scriptsize]
   // find two neighboring aggressor rows based on physical address mapping
   AGGRESSOR1, AGGRESSOR2 = find_aggressor_rows(|\textcolor{red}{{VICTIM}}|);
   // initialize the aggressor and the victim rows
   initialize(|\textcolor{red}{{VICTIM}}|, 0x55555555);
   initialize(AGGRESSOR1, AGGRESSOR2, 0xAAAAAAAA);
   // Synchronize with refresh
   for (iter = 0 ; iter < |\textcolor{red}{{NUM\_ITER}}| ; iter++):
     for (i = 0 ; i < |\textcolor{red}{{NUM\_AGGR\_ACTS}}| ; i++):
       // access multiple cache blocks in each aggressor row
       // to keep the aggressor row open longer
       for (j = 0 ; j < |\textcolor{red}{{NUM\_READS}}| ; j++): *AGGRESSOR1[j];
       for (j = 0 ; j < |\textcolor{red}{{NUM\_READS}}| ; j++): *AGGRESSOR2[j];
       // flush the cache blocks of each aggressor row
       for (j = 0 ; j < |\textcolor{red}{{NUM\_READS}}| ; j++):
         clflushopt (AGGRESSOR1[j]); 
         clflushopt (AGGRESSOR2[j]);
       mfence ();
     activate_dummy_rows();
   record_bitflips[|\textcolor{red}{{VICTIM}}|] = check_bitflips(|\textcolor{red}{{VICTIM}}|);
\end{lstlisting}

{In the extended version of our paper~\cite{rowpress-arxiv}, we investigate the effect of changing the \emph{program order} of the accesses and cache flushes to the cache blocks in the aggressor row (lines 11-16 in Algorithm~\ref{alg:rowpress-program2}). In Algorithm~\ref{alg:rowpress-program2}, every access to a cache block in the aggressor row is immediately followed by a \texttt{clflushopt} instruction (in program order), unlike in Algorithm~\ref{alg:rowpress-program} where all the \texttt{clflushopt} instructions are after all the accesses to the cache blocks in the aggressor row (in program order).}

\renewcommand{\lstlistingname}{Algorithm} %
\begin{lstlisting}[caption={RowPress demonstration program with a different program order of accesses to and flushes of cache blocks in the aggressor row.},captionpos=b,label={alg:rowpress-program2}, basicstyle=\scriptsize]
   // find two neighboring aggressor rows based on physical address mapping
   AGGRESSOR1, AGGRESSOR2 = find_aggressor_rows(|\textcolor{red}{{VICTIM}}|);
   // initialize the aggressor and the victim rows
   initialize(|\textcolor{red}{{VICTIM}}|, 0x55555555);
   initialize(AGGRESSOR1, AGGRESSOR2, 0xAAAAAAAA);
   // Synchronize with refresh
   for (iter = 0 ; iter < |\textcolor{red}{{NUM\_ITER}}| ; iter++):
     for (i = 0 ; i < |\textcolor{red}{{NUM\_AGGR\_ACTS}}| ; i++):
       // access multiple cache blocks in each aggressor row
       // to keep the aggressor row open longer
       for (j = 0 ; j < |\textcolor{red}{{NUM\_READS}}| ; j++): 
         *AGGRESSOR1[j];
         clflushopt (AGGRESSOR1[j]); 
       for (j = 0 ; j < |\textcolor{red}{{NUM\_READS}}| ; j++): 
         *AGGRESSOR2[j];
         clflushopt (AGGRESSOR2[j]);
       mfence ();
     activate_dummy_rows();
   record_bitflips[|\textcolor{red}{{VICTIM}}|] = check_bitflips(|\textcolor{red}{{VICTIM}}|);
\end{lstlisting}

{We run both programs on {our evaluated real} system {using} 1500 arbitrarily selected victim rows. {Figure~\ref{fig:real} shows the total number of bitflips {we observe} (left) and the number of rows with bitflips (right) \om{using} both Algorithm 1 (light blue bars) and Algorithm 2 (dark blue bars). The top, middle, and bottom plots show the numbers of accesses to the aggressor row per iteration being 4, 3, and 2, respectively (i.e., \texttt{NUM\_AGGR\_ACT} = 4,3,2).} The x-axis shows the {{numbers} of cache blocks read for every access to the aggressor row} (i.e., {\texttt{NUM\_READS}}).} {The leftmost bar in {each graph} shows the number of {\emph{conventional RowHammer-induced}} bitflips, {where we read \emph{only a single} cache block {per aggressor row activation} (i.e., \texttt{NUM\_READS} $=$ 1), {such that the aggressor row is kept open for a short time}}. {Remaining} bars {in each graph}  show results for RowPress-induced bitflips (with {an} increasing number of cache block reads {from left to right}, such that the {aggressor} row is kept open for an increasing amount of time).}}

{We observe that 1) both programs induce bitflips {much more effectively} {on the evaluated system by} leveraging RowPress {compared to RowHammer}, and 2) the access-flush program order in Algorithm 2 {greatly} increases the number of RowPress bitflips compared to Algorithm 1 {(by up to 98.79$\times$)}. For example, when \texttt{NUM\_AGGR\_ACT}$=$2 and $3$, RowHammer (i.e., \texttt{NUM\_READ}$=$1) does \emph{not} induce any bitflip, {RowPress induces up to 4190 bitflips in 542 DRAM rows when \texttt{NUM\_READ}$=$32 and \texttt{NUM\_AGGR\_ACT}$=$3 (Algorithm 2). When \texttt{NUM\_AGGR\_ACT}$=$4, while RowHammer induces {only} 4 bitflips in {only} 4 DRAM rows, RowPress induces up to 2371 bitflips in 429 DRAM rows (\texttt{NUM\_READ}$=$32, Algorithm 2).}}

\take{{A user-level program is much more effective at inducing bitflips on a real system by leveraging RowPress compared to RowHammer.}}

%% file: 05_mitigation.tex
\section{Mitigating RowPress}
Our characterization results and proof-of-concept real system demonstration suggest that DRAM-based systems need to take RowPress into account to {maintain the fundamental security{/safety/reliability} property of memory isolation} {and achieve robust operation}. Based on our findings, we discuss and evaluate the implications of RowPress {on} existing {{read-disturb} mitigation} mechanisms that consider {\emph{only}} RowHammer. {We propose a methodology to adapt {RowHammer} mitigation techniques} to {also} mitigate RowPress with low \emph{additional} performance overhead by both 1) limiting the \emph{maximum row-open time}, and 2) configuring the RowHammer defense to account for the {RowPress-induced} reduction in \gls{acmin}. {By applying this methodology, {our adapted mitigation techniques} 1) still mitigate RowHammer because it is more aggressive at mitigating RowHammer than the original mitigation, and 2) also mitigate RowPress because the limited maximum row-open time bounds the reduction in \gls{acmin} that the mitigation technique {needs} to keep track of.} 

{Using Ramulator~\cite{kim2016ramulator}, a cycle-accurate DRAM simulator, }{we {experimentally} demonstrate that by applying our proposed methodology to {two major {prior} techniques (}PARA~\cite{kim2014flipping} and Graphene~\cite{park2020graphene}), we can mitigate both RowHammer and RowPress with {an average (maximum)} \emph{additional} slowdown of only {$3.6\%$ ($13.1\%$)} and {$-0.63\%$ ($4.6\%$)}, respectively.{The reason for the small negative slowdowns (i.e., speedups) is that limiting the maximum row open time improve fairness between cores in a way that increases weighted speedups.}} {Our ISCA 2023 paper~\cite{luo2023rowpress} and its extended version~\cite{rowpress-arxiv} 1) provide a more detailed evaluation and analysis of {our} proposed methodology, and 2) examine and discuss three other potential mitigations {techniques for} RowPress.}

%% file: 06_significance.tex
\section{Practical Industry Impact}
We believe that our results, observations, and takeaways from our comprehensive real-device characterization, real system demonstration, and proposed mitigation of RowPress are critical for the industry to design fundamentally safe, secure, and reliable DRAM-based memory systems. RowPress opens up new a new attack surface on the memory system that could compromise the system, posing new challenges to DRAM manufacturers and system designers. We believe that existing read disturbance mitigation techniques that only {consider} RowHammer need to be adapted to also mitigate RowPress. System designers need to re-examine {adaptive row buffer management policies} in the memory controller to prevent them from being abused to conduct RowPress attacks. Future DRAM standards should also formalize and incorporate a DRAM chip's vulnerability to RowPress. Our work has received extensive media coverage~\cite{rowpressars} and was discussed heavily among technologists and industry professionals~\cite{rowpressrwt}.

\section{Research Impact}
We believe that our paper {can} inspire and enable researchers across the computing stack. Some example directions include 1) craft new attacks that leverages RowPress to break memory isolation, 2) design new mitigation techniques that better protects DRAM from read disturbance at low cost, and 3) investigate the silicon-level mechanisms of RowPress to build a {more comprehensive and fundamental} understanding of read disturbance phenomenon in DRAM. Our paper clearly shows that further research is necessary for building fundamentally safe, secure, and reliable DRAM-based memory systems. We believe and hope that our novel results, observations, and takeaways will inspire future research in both academia and industry for crafting more effective read-disturb attacks and designing more effective and efficient mitigation techniques.

%% file: 07_related_works.tex
\section{Sidebar: Related Works on DRAM Bitflips}
{To our knowledge, our ISCA 2023 paper is the first work to experimentally demonstrate and characterize RowPress and its \emph{widespread} existence commodity off-the-shelf DRAM chips. This sidebar gives an overview of the related works.}

\noindent{{\textbf{Real Chip Characterization of RowHammer.}}} {Two major prior works on experimental real DRAM chip characterization of RowHammer (\citesidebar{kim2014flipping} and ~\citesidebar{kim2020revisiting}) do not study the effect of increasing \DRAMTIMING{tAggON}. One recent RowHammer characterization work~\citesidebar{orosa2021deeper} and three device-level studies~(\citesidebar{park2016experiments}, ~\citesidebar{yang2019trap}, and ~\citesidebar{zhou2023double}) provide preliminary results on how increasing \DRAMTIMING{AggON} {\em by small amounts} affects RowHammer bitflips. These works {treat this phenomenon the same as RowHammer and} do \emph{not} {identify and demonstrate a second DRAM read-disturb phenomenon that is \emph{different} from RowHammer, while our paper clearly demonstrates the difference between RowHammer and RowPress. We refer interested readers to two overview papers(~\citesidebar{mutlu2019rowhammer} and~\citesidebar{mutlu2022fundamentally}) for more comprehensive overviews of different kinds of works on RowHammer.}}

\noindent{{\textbf{One-Location Hammering.}}} {A prior work on exploiting RowHammer~\citesidebar{gruss2018another} proposes a single-sided RowHammer technique called ``One-Location Hammering'' that ``continuously {re-opens} the same DRAM row.'' However, it is unclear whether {the} bitflips {this work observes} are caused by increased \DRAMTIMING{AggON} or conventional single-sided RowHammer.}

\noindent{{\textbf{Perspective From Industry.}}} {Two patents from {Micron}~\citesidebar{ito2017Apparatus}, and~\citesidebar{Wolff2018wordline} {very briefly mention} a ``RAS Clobber'' effect similar to RowPress. However, they do \emph{not} provide any evaluation, analysis or demonstration of this effect or clearly distinguish this effect from RowHammer. A paper from Samsung placed on arXiv~\citesidebar{hong2023dsac}
while our ISCA 2023 paper has been under review identifies a ``Passing Gate Effect'' similar to RowPress, but they do not provide any detailed explanations or real DRAM chip characterization results.}

{We call for future works to study both 1) the high-level implications of RowHammer and RowPress on designing robust computing systems, and 2) the low-level failure mechanisms of RowPress and DRAM read disturbance in general.}

\bibliographystylesidebar{IEEEtran} %
\bibliographysidebar{refs}

%% file: main.bbl
\begin{thebibliography}{10}
\providecommand{\url}[1]{#1}
\csname url@samestyle\endcsname
\providecommand{\newblock}{\relax}
\providecommand{\bibinfo}[2]{#2}
\providecommand{\BIBentrySTDinterwordspacing}{\spaceskip=0pt\relax}
\providecommand{\BIBentryALTinterwordstretchfactor}{4}
\providecommand{\BIBentryALTinterwordspacing}{\spaceskip=\fontdimen2\font plus
\BIBentryALTinterwordstretchfactor\fontdimen3\font minus \fontdimen4\font\relax}
\providecommand{\BIBforeignlanguage}[2]{{%
\expandafter\ifx\csname l@#1\endcsname\relax
\typeout{** WARNING: IEEEtran.bst: No hyphenation pattern has been}%
\typeout{** loaded for the language `#1'. Using the pattern for}%
\typeout{** the default language instead.}%
\else
\language=\csname l@#1\endcsname
\fi
#2}}
\providecommand{\BIBdecl}{\relax}
\BIBdecl

\bibitem{kim2014flipping}
Y.~{Kim} \emph{et~al.}, ``{Flipping Bits in Memory Without Accessing Them: An Experimental Study of DRAM Disturbance Errors},'' in \emph{ISCA}, 2014.

\bibitem{mutlu2019rowhammer}
O.~Mutlu \emph{et~al.}, ``{RowHammer: A Retrospective},'' \emph{IEEE TCAD}, 2019.

\bibitem{luo2023rowpress}
H.~Luo \emph{et~al.}, ``{RowPress: Amplifying Read Disturbance in Modern DRAM Chips},'' in \emph{ISCA}, 2023.

\bibitem{jedec2017ddr4}
{JEDEC}, \emph{{JESD79-4C: DDR4 SDRAM Standard}}, 2020.

\bibitem{rowpress-arxiv}
H.~Luo \emph{et~al.}, ``{RowPress: Amplifying Read Disturbance in Modern DRAM Chips},'' arXiv:2306.17061, 2023.

\bibitem{olgun2022drambender}
A.~Olgun \emph{et~al.}, ``{DRAM Bender: An Extensible and Versatile FPGA-based Infrastructure to Easily Test State-of-the-art DRAM Chips},'' \emph{IEEE TCAD}, 2023.

\bibitem{liu2013experimental}
J.~Liu \emph{et~al.}, ``{An Experimental Study of Data Retention Behavior in Modern DRAM Devices},'' in \emph{ISCA}, 2013.

\bibitem{orosa2021deeper}
L.~Orosa \emph{et~al.}, ``{A Deeper Look into RowHammer's Sensitivities: Experimental Analysis of Real DRAM Chips and Implications on Future Attacks and Defenses},'' in \emph{MICRO}, 2021.

\bibitem{hassan2021utrr}
H.~Hassan \emph{et~al.}, ``{Uncovering in-DRAM RowHammer Protection Mechanisms: A New Methodology, Custom RowHammer Patterns, and Implications},'' in \emph{MICRO}, 2021.

\bibitem{frigo2020trrespass}
P.~Frigo \emph{et~al.}, ``{TRRespass: Exploiting the Many Sides of Target Row Refresh},'' in \emph{{S\&P}}, 2020.

\bibitem{kim2016ramulator}
Y.~Kim \emph{et~al.}, ``{Ramulator: A Fast and Extensible DRAM Simulator},'' \emph{CAL}, 2016.

\bibitem{park2020graphene}
Y.~Park \emph{et~al.}, ``{Graphene: Strong yet Lightweight Row Hammer Protection},'' in \emph{MICRO}, 2020.

\bibitem{rowpressars}
{ARS Technica}, ``{There’s a new way to flip bits in DRAM, and it works against the latest defenses},'' \url{https://arstechnica.com/security/2023/10/theres-a-new-way-to-flip-bits-in-dram-and-it-works-against-the-latest-defenses/}.

\bibitem{rowpressrwt}
{Real World Tech}, ``{RowPress},'' \url{https://www.realworldtech.com/forum/?threadid=212524}.

\end{thebibliography}


\begin{thebibliography}{10}
\providecommand{\url}[1]{#1}
\csname url@samestyle\endcsname
\providecommand{\newblock}{\relax}
\providecommand{\bibinfo}[2]{#2}
\providecommand{\BIBentrySTDinterwordspacing}{\spaceskip=0pt\relax}
\providecommand{\BIBentryALTinterwordstretchfactor}{4}
\providecommand{\BIBentryALTinterwordspacing}{\spaceskip=\fontdimen2\font plus
\BIBentryALTinterwordstretchfactor\fontdimen3\font minus \fontdimen4\font\relax}
\providecommand{\BIBforeignlanguage}[2]{{%
\expandafter\ifx\csname l@#1\endcsname\relax
\typeout{** WARNING: IEEEtran.bst: No hyphenation pattern has been}%
\typeout{** loaded for the language `#1'. Using the pattern for}%
\typeout{** the default language instead.}%
\else
\language=\csname l@#1\endcsname
\fi
#2}}
\providecommand{\BIBdecl}{\relax}
\BIBdecl

\bibitem{kim2014flipping}
Y.~{Kim} \emph{et~al.}, ``{Flipping Bits in Memory Without Accessing Them: An Experimental Study of DRAM Disturbance Errors},'' in \emph{ISCA}, 2014.

\bibitem{kim2020revisiting}
J.~S. Kim \emph{et~al.}, ``{Revisiting RowHammer: An Experimental Analysis of Modern Devices and Mitigation Techniques},'' in \emph{ISCA}, 2020.

\bibitem{orosa2021deeper}
L.~Orosa \emph{et~al.}, ``{A Deeper Look into RowHammer's Sensitivities: Experimental Analysis of Real DRAM Chips and Implications on Future Attacks and Defenses},'' in \emph{MICRO}, 2021.

\bibitem{park2016experiments}
K.~Park \emph{et~al.}, ``{Experiments and Root Cause Analysis for Active-Precharge Hammering Fault in DDR3 SDRAM under 3xnm Technology},'' \emph{Microelectronics Reliability}, 2016.

\bibitem{yang2019trap}
T.~Yang \emph{et~al.}, ``{Trap-Assisted DRAM Row Hammer Effect},'' \emph{EDL}, 2019.

\bibitem{zhou2023double}
L.~Zhou \emph{et~al.}, ``Double-sided row hammer effect in sub-20 nm dram: Physical mechanism, key features and mitigation,'' in \emph{IRPS}, 2023.

\bibitem{mutlu2019rowhammer}
O.~Mutlu \emph{et~al.}, ``{RowHammer: A Retrospective},'' \emph{IEEE TCAD}, 2019.

\bibitem{mutlu2022fundamentally}
O.~Mutlu \emph{et~al.}, ``{Fundamentally Understanding and Solving RowHammer},'' in \emph{ASP-DAC}, 2023.

\bibitem{gruss2018another}
D.~Gruss \emph{et~al.}, ``{Another Flip in the Wall of Rowhammer Defenses},'' in \emph{S\&P}, 2018.

\bibitem{ito2017Apparatus}
Y.~Ito \emph{et~al.}, ``{Apparatus and Methods for Refreshing Memory},'' {U.S.}\ Patent 11062754B2, 2019.

\bibitem{Wolff2018wordline}
G.~D. Wolff, ``{Word Line Cache Mode},'' {U.S.}\ Patent 10366733B1, 2019.

\bibitem{hong2023dsac}
S.~Hong \emph{et~al.}, ``{DSAC: Low-Cost Rowhammer Mitigation Using In-DRAM Stochastic and Approximate Counting Algorithm},'' arXiv:2302.03591, 2023.

\end{thebibliography}
